\documentstyle[12pt,a4]{article}

\def\bea{\begin{eqnarray}}
\def\eea{\end{eqnarray}}
\def\be{\begin{equation}}
\def\ee{\end{equation}}

\def\nn{\nonumber}
\newcommand{\sect}[1]{\setcounter{equation}{0} \section{#1}}

\newcommand{\NPB}[3]{{\it Nucl.\ Phys.} {\bf B#1} (#2) #3}

\newcommand{\PLB}[3]{{\it Phys.\ Lett.} {\bf B#1} (#2) #3}
\newcommand{\JHEP}[3]{{\it JHEP} {\bf #1} (#2) #3}
\newcommand\eqn[1]{(\ref{#1})}
\newcommand{\ft}[2]{{\textstyle\frac{#1}{#2}}}

\def\vep{\varepsilon}

\renewcommand{\a}{\alpha}
\renewcommand{\b}{\beta}
\renewcommand{\c}{\gamma}
\renewcommand{\d}{\delta}
\newcommand{\pa}{\partial}
\newcommand{\g}{\gamma} \newcommand{\G}{\Gamma}
\newcommand{\e}{\epsilon}

\renewcommand{\l}{\lambda}
\newcommand{\m}{\mu}
\newcommand{\n}{\nu}

\newcommand{\s}{\sigma}

\renewcommand{\o}{\omega}

\newcommand\text[1]{\rm #1}

\newcommand\km{\kappa_-}

\begin{document}

\thispagestyle{empty}
\begin{flushright}
{\sc\footnotesize hep-th/9803209}\\[2mm]
{\sc THU-98/13}\\
{\sc NIKHEF 98-006}\\
\end{flushright}
\vspace{1cm}
\setcounter{footnote}{0}
\begin{center}
{\Large{\bf Superspace Geometry for Supermembrane Backgrounds}
    }\\[14mm]
{\sc Bernard de Wit${}^{\rm a,c}$, Kasper Peeters${}^{\rm b}$ and
Jan Plefka${}^{\rm b}$}\\[7mm]
{${}^{\rm a}$ \it 
Institute for Theoretical Physics, Utrecht University\\
Princetonplein 5, 3508 TA Utrecht, The Netherlands}\\
{\tt bdewit@fys.ruu.nl}\\[5mm]
{${}^{\rm b}$\it NIKHEF, P.O. Box 41882, 1009 DB Amsterdam,\\
The Netherlands}\\
{\tt t16@nikhef.nl, plefka@nikhef.nl}\\[5mm]
{${}^{\rm c}$ \it Institute for Theoretical Physics, University
    of California\\ 
Santa Barbara, CA 93106, USA}\\[20mm]

{\sc Abstract}\\[2mm]
\end{center}
We construct part of the superspace vielbein and tensor gauge field in
terms of the component fields of 
11-dimensional on-shell supergravity. The result can be utilized
to describe supermembranes and corresponding matrix
models for Dirichlet particles in nontrivial supergravity
backgrounds to second order in 
anticommuting coordinates. We exhibit the
$\kappa$-invariance of the corresponding supermembrane action,
which at this order holds for unrestricted supergravity backgrounds,
the supersymmetry covariance and the resulting surface terms in
the action.

\vfill
\leftline{{\sc March 1998}}

\newpage
\setcounter{page}{1}
\sect{Introduction}
{From} their very beginning supermembranes  \cite{BST} have 
been studied in connection with 11-dimensional supergravity \cite{CJS}.
In 11 spacetime dimensions the supermembrane can consistently
couple to a superspace background that satisfies a number of
constraints which are equivalent to the supergravity equations
of motion. In principle the supermembrane action
exists in $4,5,7$ and 11 dimensions, 
analogously to the Green-Schwarz superstring
\cite{GS84} which is classically consistent in $3,4,6$ and 10
dimensions.  But interest has focused on the 11-dimensional
supermembrane in the hope of providing a
quantum-mechanically consistent extension of supergravity in the
highest possible spacetime dimension where local supersymmetry
can exist, just as the superstring
defines such an extension for 10-dimensional
supergravities. In this context it was expected that the massless
states of the supermembrane would correspond to those of 11-dimensional
supergravity. However,  unlike the superstring states, the
supermembrane states turn out to have a continuous mass
spectrum \cite{DWLN}, which makes the possible existence of massless
states much more subtle to prove or disprove
\cite{DWHN,DWN,FH}. These
features posed an obstacle to further developments in
supermembrane theory. 

Interest in supermembranes was rekindled by the realization that
11-dimen\-sion\-al supergravity does have its role to play as the
long-distance approximation to M-theory
\cite{HT,Townsend,witten3,horvw,Townsend2}.  This theory is the
conjectured framework for unifying  
all five superstring theories and 11-dimensional supergravity. It
turns out that supermembranes, M-theory and super-matrix-models are all
intricately related. This is seen, for instance, from the result
that the light-cone formulation of the 
supermembrane  in flat backgrounds leads to a supersymmetric 
U($N$) gauge quantum-mechanical model in the large-$N$
limit \cite{DWHN}. 
This model \cite{CH}, now termed `matrix theory', has been conjectured
to capture all the degrees of freedom of M-theory
\cite{BFSS}. Furthermore 
there is evidence meanwhile that the supermembrane has massless states
\cite{mboundstates},
which will presumably correspond to the states of 11-dimensional
supergravity, although proper asymptotic states do not exist. The
existence of such states was foreseen on the basis of identifying
the 
Kaluza-Klein states of M-theory compactified on $S^1$ with the
Dirichlet particles and their bound states in type-IIA string theory.

{From} this viewpoint it is a natural question to consider the
supermembrane in curved backgrounds
associated with 11-dimensional supergravity, which is the subject
of this paper. Such backgrounds
consist of a nontrivial metric, a three-index gauge
field and a gravitino field. This provides us with an
action that transforms as a scalar under the combined (local) 
supersymmetry transformations of the background fields and the
supermembrane embedding coordinates. Here it is important to
realize that the supersymmetry transformations of the embedding
coordinates will themselves depend on the background. When
the background is supersymmetric, then the action will be
supersymmetric as well. 
In the light-cone formulation this model will lead to models
invariant under area-preserving diffeomorphisms, which in
certain situations can be approximated by matrix models 
in curved backgrounds. The area-preserving diffeomorphisms
are then replaced by a finite group, such as U$(N)$, but 
target-space diffeomorphisms are no longer manifestly
realized. Matrix 
models in curved space have already been studied in 
\cite{oldBG}. Recently toroidal compactifications of matrix
theory were considered in which the three-form gauge field of
11-dimensional gravity plays a crucial role \cite{newBG}. These
compactifications exhibit interesting features in which the
noncommutative torus appears as a new solution to compactified matrix
theory. The bosonic coupling of the membrane to the
three-form gauge field will be discussed in this paper. A summary
of this part of our results was presented earlier in
\cite{DWPP}. We should also point out that 
classical supermembrane solutions in nontrivial 
backgrounds have been discussed before, see, e.g. \cite{BDPS}. 

The approach followed in this paper for constructing the
supermembrane action is in principle straightforward and starts
from the superspace formulation presented in \cite{BST}. The
background is then characterized by the superspace vielbein and
antisymmetric tensor field. For practical calculations we like to
have a formulation in terms of the on-shell supergravity
fields. Therefore, we need to cast the component fields into
superspace, which can be done by a method sometimes referred to  
as `gauge completion' \cite{AN,BWZ,BGRS,FVN}. For 11-dimensional supergravity
the first steps of this procedure have been carried out long ago
\cite{CF}, but unfortunately only to first order in anticommuting
coordinates $\theta$. While this suffices to identify the on-shell
formulation of supergravity in superspace (see also
ref. \cite{BrinkHowe}), it is not sufficient for studying 
supermembrane interactions with the background. In this paper we
therefore extend the analysis to higher 
order in $\theta$ and are thus able to write down the  
supermembrane action in a nontrivial on-shell supergravity
background up to second order in $\theta$. 

At that point there is an important consistency check, namely
that the action is invariant under an additional local fermionic
$\kappa$-symmetry. As alluded to above,  this invariance holds provided
the background fields obey the
equations of motion of 11-dimensional supergravity
\cite{BST}. However, at second order in $\theta$ this restriction
is not yet required and our results are shown to preserve the
invariance. In this paper we concentrate on the superspace
features and we will be brief on the light-cone
formulation of the supermembrane in the supergravity
background. We intend to return to a full discussion of the
latter in a forthcoming publication \cite{future}. 
 
We have organized our paper as follows. In section 2 our
supergravity notations and conventions are established and 
as a test case the light-cone formulation of the 
bosonic membrane in curved backgrounds is studied. 
In section 3 we set the stage for an iterative
computation of the component-field content of the superfields
and superparameters 
in $\theta$, which is then taken to second order in section 4. 
All results obtained in sections 3 and 4 are collected in subsection
4.3. In section 5 we turn to the explicit form of the
supermembrane action coupled to background fields up to second order
in $\theta$ and prove its $\kappa$-symmetry. We also verify the
manifest covariance under supersymmetry, which is only a
consistency check, and determine the surface terms that follow
from $\kappa$-symmetry and supersymmetry. Finally, we discuss
possible implications and applications of our result in section~6.

\sect{Preliminaries}
\setcounter{equation}{0}
In this paper we consider superspace backgrounds that correspond to 
11-di\-men\-sion\-al supergravity. The use of certain standard conventions 
for the supermembrane will force us to employ specific and
somewhat unconventional normalizations for the 
supergravity component fields. The first subsection is therefore
devoted to a brief summary of 11-dimensional supergravity and
will establish our notation. In the second subsection we review
the supermembrane theory in superspace and indicate the effects
of a nontrivial background in its bosonic truncation. 

\subsection{Supergravity in 11 dimensions}
Supergravity in 11 spacetime dimensions is based on an ``elfbein''
field $e_\m{}^{\!r}$, a Majorana gravitino field $\psi_\m$  
and a 3-rank antisymmetric gauge field $C_{\mu\nu\rho}$.  Its 
Lagrangian\footnote{%
  Gamma matrices satisfy $\{\G^r,\G^s\} = 2\eta_{rs}$, where
  $\eta_{rs}$is the tangent-space metric $\eta_{rs}= {\rm
  diag}(-,+,\cdots,+)$. Gamma matrices with multiple indices denote
  antisymetrized products with unit strength. In particular
  $\G^{ \m_1\m_2\cdots \m_{11}} = {\bf 1}\,\varepsilon^{
  \m_1\m_2\cdots \m_{11}}$. The Dirac conjugate is defined by 
  $\bar\psi = i \psi^\dagger\Gamma^0$ for a generic spinor $\psi$. } %
can be written as follows \cite{CJS},
\bea\label{D11-lagrangian}
{\cal L}&\!\!=\!\!&  -\ft12 e\, 
R(e,\omega)  -2 e\,\bar\psi_\m\G^{\m\n\rho}D_\n[\ft12(\omega
+\hat\omega)]\psi_\rho -\ft1{96}e\, (F_{\m\n\rho\s})^2 \nonumber\\   
&& - \ft1{2\cdot 12^4}  \, \varepsilon^{\m_1\cdots \m_{11}} 
\,F_{\m_1\m_2\m_3\m_4} \,F_{\m_5\m_6\m_7\m_8} \,
C_{\m_9\m_{10}\m_{11}} \nonumber \\
&&- \ft1{96} e \,\Big(\bar\psi_\l \G^{\m\n\rho\s\l\tau} 
\psi_\tau + 12 \,\bar\psi^\m  \G^{\n\rho} \psi^\s\Big) 
  (F + \hat F)_{\m\n\rho\s} \, , 
\eea
where $e= \det e_\m{}^{r}$, 
$\omega_\m{}^{\!rs}$ denotes the spin connection and
$F_{\m\n\rho\s}$ the field strength of the antisymmetric
tensor. The caret denotes that they have been made covariant with
respect to local supersymmetry. We present the corresponding 
definitions in a sequel. The derivative $D_\mu(\omega)$ is
covariant with respect to local Lorentz transformations, 
\be
D_\m(\omega)\,\e = \Big(\pa_\m -\ft14\omega_\m{}^{rs}
\G_{rs}\Big) \e\, . 
\ee
The supersymmetry transformations are equal to 
\begin{eqnarray}
\delta e_\mu{}^r &=& 2\,\bar\epsilon\, \Gamma^r \psi_\mu \, ,
\nonumber\\
\delta \psi_\mu  &=& D_\mu(\hat\omega) \epsilon + 
T_\mu{}^{\nu\rho\sigma\kappa} \epsilon \,\hat 
F_{\nu\rho\sigma\kappa} \, , \nonumber\\
\delta C_{\mu\nu\rho} &=& -6 \,\bar\epsilon \,\Gamma_{[\mu\nu}
\psi_{\rho]}\, . 
\label{susycomp}
\end{eqnarray}
with 
\begin{equation}
T_\m{}^{\! \nu\rho\sigma\kappa} = \ft{1}{288}
\left( \Gamma_\m{}^{\!\nu\rho\sigma\kappa} - 8\, \delta^{[\nu}_\m
\Gamma^{\rho\sigma\kappa]} \right)\, .
\end{equation}
Note that $\hat F_{\m\n\rho\s}$ is the supercovariant field
strength, 
\be
\hat F_{\mu\nu\rho\sigma} = 4\,\partial_{[\mu} C_{\nu\rho\sigma]} 
+ 12 \,\bar\psi_{[\mu}\Gamma_{\nu\rho} \psi_{\sigma]}\,,
\ee
and that the supercovariant spin connection $\hat\omega^{rs}_\m$
is the solution of the following equation,
\be
D_{[\m} (\hat\omega) \,e_{\n]}^{\,r} -  \bar \psi_\m 
\G^r\psi_\n =0\, .
\label{torsion}
\ee
The left-hand side of this equation is the supercovariant torsion
tensor.  

The Lagrangian \eqn{D11-lagrangian} is derived in the context of the 
so-called ``1.5-order'' formalism, in which the spin connection is 
defined as a dependent field determined by its (algebraic) 
equation of motion, whereas its supersymmetry variation in the 
action is treated as if it were an independent field 
\cite{1.5-order}.  Furthermore we note the presence  
of a Chern-Simons-like term $F\wedge F \wedge C$
in the Lagrangian. Under tensor gauge transformations, 
\be
\d_CC_{\m\n\rho} = 3\,\pa_{[\m}\xi_{\n\rho]}\,,
\ee
the corresponding action is thus only invariant up to surface terms.  

We have the following bosonic field equations and Bianchi identities, 
\bea\label{D11-field-eq}
R_{\m\n} &=& \ft1{144}\, g_{\m\n} \,
F_{\rho\s\l\tau}F^{\rho\s\l\tau} -\ft1{12} 
F_{\m\rho\s\l}\,F_\n{}^{\!\rho\s\l}\,, \nonumber\\ 
\pa_{\m}\Big(e \,  F^{\m\n\rho\s}\Big) &=& \ft1{1152} \, 
\varepsilon^{\n\rho\sigma\m_1\ldots\m_{8}} 
F_{\m_1\m_2\m_3\m_4}\,F_{\m_5\m_6\m_{7}\m_{8}}\,, \nonumber \\
\pa_{[\m}F_{\n\rho\s\l]}&=&0\,,
\eea
which no longer depend explicitly on the antisymmetric gauge field. 
An alternative form of the second equation is \cite{page}
\be\label{D11-field-eq2}
\pa_{[\m_1}H_{\m_2\ldots\m_8]} =0\,,
\ee
where $H_{\m_1\ldots\m_7}$ is the dual field strength,
\be
H_{\m_1\ldots\m_7} = 
{1\over 7!}e\, 
\varepsilon_{\m_1\ldots\m_{11}} 
F^{\m_8\m_9\m_{10}\m_{11}} 
-\ft1{12}  \, F_{[\m_1\m_2\m_3\m_4}\,C_{\m_5\m_6\m_7]}\,.
\ee
When the third equation of \eqn{D11-field-eq} and 
\eqn{D11-field-eq2} receive contributions from certain source
terms on the right-hand side, then the corresponding charges 
can be associated with the
`flux'-integral of  $H_{\m_1\ldots \m_7}$ 
and $F_{\m_1\m_2\m_3\m_4}$ over the boundary of an 8- and a 5-dimensional 
spatial volume, respectively. This volume is transverse to a  
$p=2$ and $p=5$ brane configuration, and the corresponding
charges are 2- and  5-rank Lorentz tensors. For solutions of 
11-dimensional supergravity that contribute to these charges, see
e.g. \cite{sugramem,sugrafive,solirev,Townsend2}. 

It is straightforward to evaluate the supersymmetry algebra on 
these fields. The commutator of two supersymmetry transformations 
yields a general-coordi\-nate transformation, a supersymmetry transformation,
a local Lorentz transformation, and a gauge 
transformation associated with the tensor gauge field,
\be
[\d(\e_1),\d(\e_2) ] = \d_{\rm gct} (\xi^\m) + \d(\e_3)
+ \d_L(\l^{rs}) + \d_C(\xi_{\m\n}) \,.
\ee
The parameters of the transformations on the right-hand side 
are given by
\bea
\xi^\m&\!=\!& 2 \,\bar\e_2\G^\m\e_1 \,,\nonumber\\[1.1mm] 
\e_3&\!=\!& -\xi^\m \psi_\m \,, \nonumber\\[.8mm]
\l^{rs} &\!=\!& -\xi^\m\, \hat \omega_\m{}^{rs} + \ft1{72}\, 
\bar\e_2\Big[ \G^{r s\mu\nu\rho\sigma}\hat F_{\mu\nu\rho\sigma} + 24\,   
\G_{\m\nu}\hat F^{rs\mu\nu}\Big]\e_1\,, \nonumber\\
\xi_{\mu\nu} &\!=\!& - \xi^\rho\, C_{\rho\mu\nu}
-2 \,\bar \e_2 \G_{\mu\nu}\e_1\,. \label{compocomm}
\eea

\subsection{Membranes in background fields}
The 11-dimensional supermembrane \cite{BST} is written in terms of 
superspace embedding coordinates 
$Z^M(\zeta)=(X^\mu(\zeta),\theta^\alpha(\zeta))$, which are functions of 
the three world-volume coordinates $\zeta^i$ ($i = 0,1,2$). It 
couples to the superspace geometry of 11-dimensional supergravity,
encoded by the supervielbein $E_M{}^A$ and the antisymmetric tensor
gauge superfield $B_{MNP}$, through the action\footnote{%
  Our notation and conventions are as follows. Tangent-space indices are
  $A=(r,a)$, whereas curved indices are denoted by
  $M=(\mu,\alpha)$. Here $r,\mu$ refer to  commuting and
  $a,\alpha$ to anticommuting coordinates. Moreover we take 
  $\epsilon_{012}=-\epsilon^{012}=1$. } %
\be
S[Z(\zeta)] =\int {\rm d}^3\zeta \;\Big[- \sqrt{-g(Z(\zeta))} - \ft16
\varepsilon^{ijk}\, \Pi_i^A \Pi_j^B\Pi_k^C \,B_{CBA}(Z(\zeta)) 
\,\Big]\,,
\label{supermem}
\ee
where $\Pi^A_i = \pa Z^M/\pa\zeta^i \; E_M{}^{\!A}$ is the pull-back of
the supervielbein to the membrane worldvolume. Here the 
induced metric equals $g_{ij} =\Pi^r_i \Pi^s_j \,\eta_{rs}$, with
$\eta_{rs}$ being the constant Lorentz-invariant metric. This action
is invariant under local fermionic $\kappa$ transformations \cite{BST},
given that certain constraints on the background fields hold, which are 
equivalent to the  equations of motion of 11-dimensional 
supergravity \eqn{D11-field-eq}.

Flat superspace is characterized by 
\bea\label{flatssquantities}
E_\mu{}^r &\!\!=\!\!& \d_\mu{}^r \, , \hspace{43mm} E_\mu{}^a =0  \, ,
\nonumber \\
E_\alpha{}^a &\!\!=\!\!& \delta_\alpha{}^a  \, ,\hspace{42.6mm} 
E_\alpha{}^r = -(\bar\theta \Gamma^r)_{\alpha} \, ,\nonumber \\
B_{\mu\n\alpha} &\!\!=\!\!& (\bar\theta\Gamma_{\m\n})_{\alpha}\,
, \hspace{32.65mm}
B_{\m\a\b} = 
(\bar\theta\Gamma_{\m\n})_{(\a}\,(\bar\theta\Gamma^\n)_{\b)} \,,\nonumber \\
 B_{\a\b\g} &\!\!=\!\!& (\bar\theta\Gamma_{\m\n})_{(\a}\, 
(\bar\theta\Gamma^\m)_{\b}\, 
(\bar\theta\Gamma^\n)_{\g)}\,,\hspace{6.2mm}  
B_{\mu\nu\rho} =0  \, .  
\eea
These quantities receive corrections in the presence of 
supergravity background fields $e_\mu{}^r$, $\psi_\mu^a$ and 
$C_{\mu\nu\rho}$, and it is the aim of this paper to determine
some of these corrections to second order in $\theta$. 

In flat  
superspace the supermembrane Lagrangian, written in 
components, reads,  
\bea
{\cal L} &=&- \,\sqrt{-g(X,\theta)} \nonumber \\
&&- \varepsilon^{ijk} \,
\overline\theta\Gamma_{\mu\nu}\partial_k\theta \Big
[{\textstyle{1\over 2}} 
\,\partial_i X^\mu (\partial_j X^\nu +
\overline\theta\Gamma^\nu\partial_j \theta) + {\textstyle{1\over 6}}
\,\overline\theta\Gamma^\mu\partial_i\theta\;
\overline\theta\Gamma^\nu\partial_j\theta \Big] \,, 
\label{action}
\eea

To elucidate the generic effects of nontrivial backgrounds for
membrane theories, let us confine ourselves for the moment 
to the purely bosonic theory and present
the light-cone formulation of the membrane in a 
background consisting of the metric $G_{\mu\nu}$ and the tensor 
gauge field $C_{\mu\nu\rho}$.
The Lagrangian density for the bosonic membrane follows directly 
{}from \eqn{supermem},  
\begin{equation}
{\cal L} = -\sqrt{-g} - \ft{1}{6}\varepsilon^{ijk} \partial_i X^\mu\,
\partial_j
X^\nu \,\partial_k X^\rho\, C_{\rho\nu\mu} \, ,
\end{equation}
where $g_{ij}= \pa_i X^\m \,\pa_j X^\n \,\eta_{\m\n}$.  
In the light-cone formulation, the coordinates are decomposed in 
the usual fashion as $(X^+,X^-,X^a)$ with  
$a=1\ldots 9$. Furthermore we use the diffeomorphisms in the 
target space to bring the metric in a convenient form~\cite{GS},
\begin{equation}
\label{metricgauge}
G_{--}=G_{a-}=0 \, .
\end{equation}
Subsequently we identify the time coordinate of the target space 
with the world-volume time, by imposing the condition  $X^+ = 
\tau$. Moreover we denote the spacesheet coordinates of the membrane
by $\sigma^r$, $r=1,2$.
Following the same steps as for the membrane in flat 
space~\cite{DWHN}, one arrives at a Hamiltonian formulation of 
the theory in terms of coordinates and momenta. These
phase-space variables are subject to a constraint, which 
takes the same form as for the membrane theory in flat space, namely,  
\begin{equation}
\phi_r = P_a \,\partial_r X^a + P_- \,\partial_r X^- \approx 0\, .
\label{constraintmem}
\end{equation}
Of course, the definition of the momenta in terms of the 
coordinates and their derivatives does involve the 
background fields, but at the end all explicit dependence 
on the background cancels out. 

The Hamiltonian now follows straightforwardly. As it turns out,
the background tensor field appears in the combinations
\bea
C_a &=& - \varepsilon^{rs} \partial_r X^- \partial_s X^b \,C_{-ab} +
\ft{1}{2}
\varepsilon^{rs}\partial_r X^b \partial_s X^c \,C_{abc} \, ,      
\nonumber \\
C_{\pm} &=& \ft{1}{2}\varepsilon^{rs}\partial_r X^a \partial_s X^b
\,C_{\pm ab}\,, \nonumber\\
C_{+-} &=& \varepsilon^{rs}\partial_r X^- \partial_s X^a\,
C_{+-a}\,.
\eea
With these definitions the total Hamiltonian takes the form
\bea
H&\!=\!& \int {\rm d}^2\sigma\, \bigg \{
\frac{G_{+-}}{P_- - C_-}\bigg[\ft{1}{2}\Big(P_a-C_a-\frac{P_--C_-}{G_{+-}}
\, G_{a+}\Big)^2+ \ft{1}{4}
(\varepsilon^{rs}\, \partial_r X^a\, \partial_s X^b )^2\bigg]\nonumber\\
&&\hspace{15mm}  -\frac{P_--C_-}{2\, G_{+-}}\, G_{++}- C_+-C_{+-}
+ c^r \phi_r \bigg \} \,. \label{BGHAM}
\eea
where we have included the Lagrange multiplier $c^r$ coupling to the
constraint \eqn{constraintmem}. Observe that transverse indices
are contracted with the metric $G_{ab}$ or its inverse.

The gauge choice $X^+=\tau$ still allows for $\tau$-dependent 
reparametrizations of the world-space coordinates
$\sigma^r$, which in turn induce transformations on the Lagrange multiplier
 $c^r$ through the Hamilton equations of motion.
In addition there remains the freedom of performing tensor gauge
transformations of the target-space three-form $C_{\mu\nu\rho}$.
In order to rewrite \eqn{BGHAM} in terms of a gauge theory of
area-preserving diffeomorphisms it is desirable to obtain a Hamiltonian
which is polynomial in momenta and coordinates.
For this the dynamics of
$P_--C_-$ needs to become trivial, i.e. $\partial_\tau (P_--C_-)=0$, 
allowing us to set it equal to some space-sheet density
$\sqrt{w(\s)}$. The residual invariance group is then constituted by the
area-preserving diffeomorphisms that leave $\sqrt w$ invariant.
The $\tau$-independence of $P_--C_-$ can be achieved by firstly
assuming that the background fields  
are $X^\pm$-independent. Secondly one uses the tensor
gauge transformations to set $C_{-ab}$ equal to a constant
antisymmetric matrix. One then has
\be
\partial_\tau (P_--C_-)\approx \partial_r\, \Bigl [
-\varepsilon^{rs}\partial_s X^a\, C_{+-a} + (P_--C_-)\, c^r\, 
\Bigr ] \,. 
\label{crdef}
\ee
We now choose a gauge such that the right-hand side of this
equation vanishes. In that case the total Hamiltonian takes the
following form, 
\bea
H&\!=\!& \int {\rm d}^2\sigma\, \bigg \{
\frac{G_{+-}}{P_- - C_-}\bigg[\ft{1}{2}\Big(P_a-C_a-\frac{P_--C_-}{G_{+-}}
\, G_{a+}\Big)^2+ \ft{1}{4}
(\varepsilon^{rs}\, \partial_r X^a\, \partial_s X^b )^2\bigg]\nonumber\\
&&\hspace{15mm}  -\frac{P_--C_-}{2\, G_{+-}}\, G_{++}- C_+\nn\\
&&\hspace{15mm} +{1\over{P_--C_-}} \Big[ \varepsilon^{rs} \pa_r
X^a\pa_s X^b \,P_a\,C_{+-b} + C_-\,C_{+-}\Big]  \bigg \} \,, \label{BGham}
\eea
where $P_--C_-\propto \sqrt w$ and $C_{-ab}$ constant. 

At this point one can impose further gauge choices and set
$G_{+-}=1$ and $C_{+-a}=0$. Taking also $C_{-ab}=0$ the
corresponding Hamiltonian was 
recast in Lagrangian form in \cite{DWPP} in
terms of a gauge theory of area-preserving diffeomorphisms. With
both $C_{+-a}$ and $C_{-ab}$ different from zero, one can go
through the same procedure.
As alluded to in the first reference of \cite{newBG}, the
Lagrangian then depends explicitly on $X^-$, a feature that we
have already exhibited earlier for the winding membrane
\cite{us}. However, in the case at hand, the $X^-$-dependence is
rather nontrivial and clearly this is an issue that deserves more study. 

With a reformulation of the membrane in background  fields as a 
gauge theory of area-preserving diffeomorphisms at one's
disposal, one may consider its regularization through a matrix
model by truncating 
the mode expansion for coordinates and momenta in the standard fashion
\cite{GoldstoneHoppe,DWHN}. This leads to a replacement of Poisson
brackets by commutators, integrals by traces and products of
commuting fields by symmetrized products of the corresponding 
matrices. At that point the original target-space covariance
is affected, as the matrix reparametrizations in terms of
symmetrized products of matrices do not possess a consistent
multiplication structure\footnote{%
  We thank J. de Boer for explaining this to us. }; %
this is just one of the underlying difficulties in the
construction of matrix models in curved space \cite{oldBG}. Recently the
antisymmetric constant matrix $C_{-ab}$ was conjectured to 
play a role for the matrix model compactification on a
noncommutative torus \cite{newBG}. It should be interesting to
see what the role is of \eqn{BGham} in this context. 
We intend to return to these issues in more detail in a
future publication \cite{future}.

\section{Superspace representation}
\setcounter{equation}{0}
In this section we introduce the method for constructing 
superspace backgrounds expressed in terms of the component fields
of 11-dimensional supergravity. Besides this we evaluate the
quantities of interest in low orders of the anticommuting
coordinates $\theta$. 
The superspace coordinates $Z^M$ are given by $Z^M=
(x^\m,\theta^\a)$. 
The superspace geometry is encoded in the supervielbein
$E_M{}^{\!A}$ and a spin-connection field $\Omega_M{}^{\!AB}$. In
what follows we will not pay much 
attention to the spin-connection, which is not an independent
field. Furthermore we have an antisymmetric 
tensor gauge field $B_{MNP}$, subject to tensor gauge transformations, 
\be
\d B_{MNP} = 3\,\partial_{[M} \Xi_{NP]}\, .
\ee 
Unless stated otherwise the derivatives with respect to $\theta$
are always left derivatives. We remind the reader that
`antisymmetric' tensors in superspace satisfy the symmetry
properties induced by those of superdifferentials, i.e. ${\rm
d}Z^M\wedge {\rm d}Z^N= (-)^{1+MN}\, {\rm
d}Z^N\wedge {\rm d}Z^M$. 

Under superspace diffeomorphisms corresponding to $Z^M\to
Z^M+\Xi^M(Z)$, the super-vielbein and tensor gauge field
transform as  
\bea
\delta E_M{}^{\!A} &=& \Xi^N \partial_N E_M{}^{\!A} + \partial_M \Xi^N
E_N{}^{\!A} \,,\nonumber \\ 
\delta B_{MNP} &=& \Xi^Q \partial_Q B_{MNP} 
+ 3\,\partial_{[M} \Xi^Q B_{\vert Q\vert NP]}\,.
\eea
Tangent-frame rotations are $Z$-dependent Lorentz
transformations that act on the vielbein according to
\be
\delta E_M{}^A =  \ft{1}{2}
(\Lambda^{rs}L_{rs})^A{}_{\!B}\, E_M{}^{\!B}\,,
\ee
where the Lorentz generators $L_{rs}$ are defined by
\be
\ft{1}{2}(\Lambda^{rs} L_{rs})^t{}_u = \Lambda^t{}_u \, ,\qquad
\ft{1}{2}(\Lambda^{rs} L_{rs})^a{}_b = \ft{1}{4}\Lambda^{rs}
(\Gamma_{rs})^a{}_b\, . 
\ee

We will not be dealing with an unrestricted superspace but one
that is subject to certain constraints and gauge
conditions. Furthermore, we will not describe an
off-shell situation 
as all superfields will be expressed entirely in terms of 
the three component fields of on-shell 11-dimensional supergravity,
the elfbein $e_\m{}^r$, the antisymmetric tensor gauge field
$C_{\m\n\rho}$ and the gravitino field $\psi_\m$. As a result of
these restrictions the residual symmetry transformations are confined to 
11-dimensional diffeomorphisms with parameters $\xi^\m(x)$, 
local Lorentz transformations with parameters $\l^{rs}(x)$, 
tensor-gauge transformations with parameters $\xi_{\m\n}(x)$ and
local supersymmetry transformations with parameters $\e(x)$. The
purpose of this paper is to derive how the superfields are
parametrized in terms of the component fields. To do this it is
necessary to also determine the form of the superspace
transformation parameters, $\Xi^M$,
$\Lambda^{rs}$ and $\Xi_{MN}$, that generate the supersymmetry
transformations. Here it is important to realize
that we are dealing with a gauge-fixed situation. For that reason
the superspace parameters depend on both the $x$-dependent
component parameters defined above as well as on the component 
fields. 
This has two consequences. First of all, local
supersymmetry transformations reside in the superspace
diffeomorphisms, the Lorentz transformations and the
tensor gauge transformations, as $\Xi^M$,
$\Lambda^{rs}$ and $\Xi_{MN}$ are all expected to contain
$\epsilon$-dependent terms. Thus, when considering 
supersymmetry variations of the various fields, one must in
principle include each of the three possible superspace
transformations. Secondly, when 
considering the supersymmetry algebra, it is crucial to
also take into account the variations of the component fields on
which the parameters $\Xi^M$,
$\Lambda^{rs}$ and $\Xi_{MN}$ will depend. 

In this section we present the formalism and derive the
expressions for the various 
superfields in terms of component fields in low orders of $\theta$. 
This will set the stage for the
evaluation of the terms of higher order in $\theta$, which is the
subject of the next section, and it will allow us to establish
the precise correspondence with the flat
superspace conventions used for the supermembrane action in the
previous subsection. The method of
casting component results into superspace has a long history and
is sometimes called `gauge completion'. For results in 4
spacetime dimensions 
we refer the reader to \cite{BGRS,FVN}, while results in 11 dimensions
in low orders of $\theta$ were presented in \cite{CF}.

There are two, somewhat complimentary, ways to obtain information
on the embedding of component fields in superspace geometry. One
is to consider the algebra of the supersymmetry
transformations as generated by the superspace transformations and
to adjust it to the supersymmetry algebra of the component
fields. This determines the 
superspace transformation parameters. The other is to compare the
transformation rules for the  
superfields with the known transformations of the component
fields. This leads to a parametrization of both the superfields
and the transformation parameters in terms
of the component fields and parameters. The evaluation proceeds
order-by-order in 
the $\theta$-coordinates, but at each level one 
encounters ambiguities which can be fixed by suitable
higher-order coordinate redefinitions and gauge choices. The first
step in this iterative procedure is the  
identification at zeroth-order in $\theta$ of some of the component 
fields and transformation parameters with corresponding
components of the superfield
quantities. The underlying assumption is that this identification
can always be implemented by choosing an appropriate gauge. An 
obvious identification is given by 
\cite{AN,BWZ,BGRS,FVN,CF},  
\bea
E_\m{}^r(x,\theta=0) &=& e_\m{}^r(x)\,,  \nonumber \\
E_\m{}^a(x,\theta=0) &=& \psi_\m{}^a(x)\,,\nonumber \\
B_{\m\n\rho}(x,\theta=0) &=& C_{\m\n\rho}(x)\,,\nonumber \\
\Xi^\m(x,\theta=0) &=& \xi^\m(x)\,, \nonumber \\
\Xi^\a(x,\theta=0) &=& \e^\a(x)\,, \nonumber \\
\Lambda^{rs}(x,\theta=0) &=& \lambda^{rs}(x)\,,\nonumber \\
\Xi_{\m\n}(x,\theta=0) &=& \xi_{\m\n}(x)\,.  \label{initial}
\eea

As explained above, the component supersymmetry transformations
with parameters $\e(x)$ are generated by a linear combination of
a superspace diffeomorphism, a local Lorentz and a tensor gauge
transformation; their corresponding parameters will be 
denoted by $\Xi^M(\e)$, $\Lambda^{rs}(\e)$ and $\Xi_{MN}(\e)$,
respectively. Given the embedding of the component fields into
the superfields, application of these specific superspace
transformations should produce the very same  transformation
rules that were defined directly at the component level. The
structure of the commutator algebra of  
unrestricted infinitesimal superspace transformations is
obvious. Two 
diffeomorphisms yield another diffeomorphism, two Lorentz
transformations yield another Lorentz transformation, according to
the Lorentz group structure, while two tensor transformations
commute. On the other hand, a diffeomorphism
and a local Lorentz transformation yield another Lorentz
transformation, and a diffeomorphism and a tensor gauge
transformation yield another gauge transformation. All other
combinations commute. 

The algebra for the restricted superspace transformations that
generate the component transformations should coincide with the
algebra derived directly for the component fields. However, we
must take into account here that the superspace transformation
parameters themselves depend on the component fields. To show the
effect of this let us restrict ourselves to the diffeomorphism
component of the supersymmetry transformation and consider the
transformation of a generic scalar superfield, 
\be 
\d(\e)\Phi = \Xi^M(\e)\, \pa_M\Phi\,.
\ee
Closure of supersymmetry now implies 
\be
\d(\e_1,\e_2)\,\Phi= [\d(\e_1),\d(\e_2)]\,\Phi =
\Xi^M(\e_1,\e_2)\,\pa_M \Phi\,, 
\ee
with 
\be
\Xi^M(\e_1,\e_2) = \Xi^N(\e_2)\,\pa_N \Xi^M(\e_1) + \d(\e_1)\,
\Xi^M(\e_2)  - (1\leftrightarrow 2) \,. \label{diffalgebra}
\ee
Here the variation $\d(\e_1,\e_2)$ represents the component
result \eqn{compocomm} of the supersymmetry commutator and
$\Xi^M(\e_1,\e_2)$ represents the part of the resulting component
transformations that are generated by superspace
diffeomorphisms. Note that we are justified in restricting 
ourselves to the superspace diffeomorphisms, because they are the
only ones that lead to a superspace diffeomorphism upon
commutation. We will consider the other superspace
transformations later. At zeroth-order in $\theta$ we compare the
expression for $\Xi^M(\e_1,\e_2)$ to the result of the component
supersymmetry algebra, taking into account the
conditions \eqn{initial}. As it turns out, this leads to the
following result,
\bea
\Xi^\m(\e) &=& \bar\theta\, \G^\m \e
+\sum_{n=0,3,4}\,\bar\theta\,\Gamma^{r_1\cdots r_n}\e\,{\cal
H}^\m_{r_1\cdots r_n} + {\cal O}(\theta^2)\,, \nonumber\\ 
\Xi^\a(\e) &=& \e^\a -\bar\theta \,\G^\m\e\,\psi_\m^\a +
\sum_{n=0,3,4}\,\bar\theta\,\Gamma^{r_1\cdots r_n}\e\,{\cal
H}^\a_{r_1\cdots r_n} +  {\cal O}(\theta^2)\,, 
\eea 
where the ${\cal H}^M_{r_1\cdots r_n}$ are undetermined
$\theta$-independent quantities. 

Subsequently one compares the supersymmetry variations at
$\theta=0$ of the supervielbein components to their variation 
under a diffeomorphism given by $\Xi^M(\e)$. In principle one has
to allow for a local Lorentz transformation here, but for
$\theta=0$ it vanishes. The comparison results in the 
following values for the supervielbein components,
\bea
E_\m{}^{\!r} &=& e_\m{}^r + 2\, \bar\theta\,\G^r \psi_\m +  {\cal
O}(\theta^2)\,, \nonumber \\[3mm]
E_\m{}^{\!a} &=& \psi_\m{}^{\!a} - \ft14\,
\hat\o_\m^{rs}\,(\G_{rs} \theta)^a + 
(T_\m{}^{\!\n\rho\s\lambda}\theta)^a\, \hat F_{\n\rho\s\lambda} +
{\cal O}(\theta^2)\,, \nonumber \\[2mm]
E_\a{}^{\!r} &=& -(\bar\theta\,\G^r)_\a +
\sum_{n=0,3,4}\,(\bar\theta\,\Gamma^{r_1\cdots r_n})_\a \,{\cal
H}^r_{r_1\cdots r_n}+  {\cal O}(\theta^2)\,, \nonumber \\ 
E_\a{}^{\!a} &=& \d^a_\a + \sum_{n=0,3,4}\,(\bar\theta\,\Gamma^{r_1\cdots
r_n})_\a \,{\cal H}^a_{r_1\cdots r_n} +  {\cal O}(\theta^2)\,. 
\eea

Let us briefly discuss these results. First of all we are dealing
with an ambiguity in the iterative procedure reflected in the
presence of the $\theta$-independent quantities ${\cal
H}^M_{r_1\cdots r_n}$. However, it turns out 
that this ambiguity can be absorbed into the definition of the
superspace coordinates, according to 
\be
Z^M \to Z^M + \ft12 \sum_{n=0,3,4}\,
\bar\theta\,\G^{r_1 \cdots r_n}\theta \, {\cal H}^M_{r_1\cdots
r_n}\,.
\ee
Hence we may set ${\cal H}^M_{r_1\cdots r_n}=0$ in what
follows. In that case 
our results for the vielbein agree with the flat-space 
expressions \eqn{flatssquantities} employed for the supermembrane in
the previous  
section (and corresponding to $\hat \omega_\m^{rs}= \hat
F_{\m\n\rho\s}=\psi_\m=0$ and $e_\m{}^{\!r} = \d_\m^r$).  

Furthermore, the fact that $E_\a{}^a=\d^a_\a$ in this
order implies that the local Lorentz transformation will be
accompanied henceforth by a corresponding diffeomorphism given by
\be
\Xi^\a(\lambda) = -\ft14 \lambda^{rs} (\G_{rs}
\theta)^\a\,. \label{thetalorentz} 
\ee
This term ensures that the various superspace components take a
covariant form with respect to the local Lorentz transformations
parametrized by $\lambda^{rs}$. In due course \eqn{thetalorentz}
will also arise at 
higher orders in $\theta$ in the gauge 
completion procedure, as the supersymmetry commutator contains
a field-dependent Lorentz transformation. A corresponding phenomenon 
does not occur for the 11-dimensional diffeomorphisms and the tensor
gauge transformations parametrized by $\xi^\m$ and $\xi_{\m\n}$,
which do not entangle with other superspace components. This is
so because the initial conditions \eqn{initial} are fully
covariant with respect to these transformations.  

Before continuing we note that the components of $\Xi^A=
\Xi^M\,E_M{}^{\!A}$ remain field-independent. One expects that
these tangent-space expressions will be supercovariant, so that
the gravitino or the spin-connection fields cannot appear
explicitly (for a discussion of this property, see
\cite{BWZ,FVN}). The field-independent 
values for these expressions are given by 
\be
\Xi^M\, E_M{}^{\!r} = 2\,\bar\theta\,\G^r\e(x)\,,\qquad 
\Xi^M\, E_M{}^{\!a} = \e^a(x)\,. \label{tangentdiffs}
\ee
The above result can be regarded to some extent as a gauge
condition. To see this, one may verify that (in this order of
$\theta$) it implies that the ambiguities encoded in ${\cal
H}^r_{r_1\cdots r_n}$ vanish.
In the next section we will confirm the validity of the first of
these relations to order $\theta^2$. The second relation, however, 
will receive 
contributions proportional to $\hat F_{rstu}$ (written with flat
indices). We will refrain from calculating these terms as they are not
directly relevant for the purpose of this paper. 

Let us now turn to the tensor field. The supersymmetry commutator
for the component fields gives rise to a field-dependent tensor gauge
transformation. Such a gauge transformation can arise because the
tensor field is subject to both superspace diffeomorphisms and 
tensor gauge transformations. The commutator of a diffeomorphism
and a tensor gauge transformation gives again a tensor gauge
transformation and this leads to the component result. Hence the
result \eqn{diffalgebra} is incomplete for the tensor field and
there is an extra tensor transformation given by 
\bea
\Xi_{MN}(\e_1,\e_2) &=& \Xi^P(\e_2)\,\pa_P\Xi_{MN}(\e_1) +
2\,\pa_{[M}\Xi^P(\e_2)\,\Xi_{\vert P\vert N]}(\e_1) + \d(\e_1)
\Xi_{MN}(\e_2)  \nonumber \\
&& - (1 \leftrightarrow 2) \,. \label{diff-tensor}
\eea
Before evaluating this equation we first note that the
transformation parameters $\Xi_{MN}$ are only defined up to terms
of the form $\pa_{[M}\Lambda_{N]}$. We can use this feature to
set all $\Xi_{MN}(x,\theta=0)$ other than
$\Xi_{\m\n}(x,\theta=0)$ to zero (for 
this one chooses the $\Lambda_M$ linear in $\theta$). 
With this simplification we compare \eqn{diff-tensor} at
$\theta=0$ to the tensor component in the supersymmetry algebra
\eqn{susycomp} and we find
\bea
\Xi_{\m\n}(\e)&=& \bar \e(C_{\m\n\rho}\,\G^\rho + \G_{\m\n} )\theta +
\sum_{n=0,3,4}  \,\bar\e\,\G^{r_1\cdots r_n}\theta \, {\cal
H}_{\m\n\,r_1\cdots r_n}+  {\cal O}(\theta^2)\,, \nonumber \\
\Xi_{\m\a}(\e) &=& 
\sum_{n=0,3,4}  \,\bar\e\,\G^{r_1\cdots r_n}\theta \, {\cal
H}_{\m\a\,r_1\cdots r_n}     +  {\cal O}(\theta^2)\,, \nonumber \\
\Xi_{\a\b}(\e) &=& 
\sum_{n=0,3,4}  \,\bar \e\,\G^{r_1\cdots r_n}\theta \, {\cal
H}_{\a\b\,r_1\cdots r_n} +  {\cal O}(\theta^2) \,.
\eea
Again there are undetermined terms characterized by
$\theta$-independent quantities ${\cal H}_{MN\,r_1\cdots r_n}$.

With this result we consider the variations of $B_{MNP}$ under a
combined superspace diffeomorphism and tensor gauge
transformation. However, first we note that we can set all the
components of $B_{MNP}(x,\theta=0)$, with the
exception of $B_{\m\n\rho}(x,\theta=0)$, to zero by suitable gauge
transformations with parameters linear in $\theta$. We then
establish the following results,
\bea
B_{\m\n\rho} &=& C_{\m\n\rho} -6\, \bar \theta
\G_{[\m\n}\psi_{\rho]}   +  {\cal O}(\theta^2) \,,\nonumber\\[3mm] 
B_{\m\n\a} &=& (\bar \theta \,\G_{\m\n})_\a  +
\sum_{n=0,3,4} \,(\bar\theta\,\G^{r_1\cdots r_n} )_\a \, {\cal
H}_{\m\n\,r_1\cdots r_n}     +  {\cal O}(\theta^2)
\,, \nonumber\\ 
B_{\m\a\b} &=& 2\sum_{n=0,3,4}  \,(\bar\theta\,\G^{r_1\cdots
r_n})_{(\a} \, {\cal H}_{\b)\m\,r_1\cdots r_n}   +  {\cal
O}(\theta^2) \, ,  \nonumber \\
B_{\a\b\c} &=& 3 \sum_{n=0,3,4}  \,\bar (\theta\,\G^{r_1\cdots
r_n})_{(\a} \, {\cal H}_{\b\gamma)\,r_1\cdots r_n}   +  {\cal
O}(\theta^2) \,. 
\eea
Subsequently we note that all the ambiguous terms proportional to
${\cal H}_{MN\,r_1\cdots r_n}$ can be removed by a gauge
transformation with parameters proportional to $\theta^2$ and
equal to
\be
\Xi_{MN}= -\ft12 \sum_{n=0,3,4} \,\bar\theta\,\G^{r_1\cdots
r_n}\theta \, {\cal H}_{MN\,r_1\cdots r_n}  \,.
\ee 
Hence we drop these terms here so that also the
results pertaining to the tensor field agree with the
flat-space values \eqn{flatssquantities} used in the previous
section. 

\section{Higher-order contributions}
\setcounter{equation}{0}
So far our results are in agreement with those of \cite{CF}. In
this section we determine the higher-order contributions and go
beyond the results reported in the literature. In higher orders a
number of new features enters, which did not play a role in the
previous section. First of all the Lorentz transformations acting
on the vielbein will now become relevant as well as the supersymmetry
variation of the fields in the transformation parameters when
evaluating the supersymmetry commutators. The 
reason why the Lorentz transformations did not enter earlier is
related to the fact that we did not consider the components of the
superspace spin connection. The
ambiguities noted in the previous section will persist, but we
will no longer exhibit their explicit form in order to keep our
expressions tractable. Nevertheless, we have convinced ourselves
that they can be gauged away in the same fashion as before. 
The presence of higher-order spinor terms unavoidably leads to
the need of Fierz reorderings, which tend to be rather
cumbersome in 11 dimensions. However, in all cases we could avoid
explicit reorderings by  making use of the well-known
identity, which holds in $4,5,7$ and 11 spacetime dimensions, 
\be
\bar\psi_{[1}\G^\n\psi_2\;\bar\psi_3\G_{\m\n}\psi_{4]} = 0\,.
\label{4spinor-id}
\ee

Below we start by deriving the higher-order expressions for the
vielbein and, in a second subsection, for the tensor
field. We will not always, as 
before, completely  exploit the supersymmetry commutator, but
sometimes move directly to the field variations and confront
their form   
with that induced by a superspace diffeomorphism combined with a 
Lorentz or with a tensor gauge transformation. In a
third subsection we present a summary of all the terms obtained.

\subsection{The vielbein at order $\theta^2$}
We start with \eqn{diffalgebra} for $M=\mu$ at order $\theta$,
where we now 
must take into account the transformation of the component fields
appearing in the superspace parameters. Using the lower-order
results obtained previously and the value for $\e_3$ given in
\eqn{compocomm}, one can integrate the equation  and
obtains
\be
\Xi^\m(\e)\Big\vert_{\theta^2} = -\bar \theta\,
\G^\n\e\;\bar\theta\,\G^\m\psi_\n\,. \label{xi-mu-2} 
\ee
This result is not unique and defined up up to an expression 
\be
{\cal H}^\m_{\a\b\gamma} \,\e^\a\,\theta^\b\theta^\g \,,\label{ambi} 
\ee
with $\cal H$ a tensor antisymmetric in $[\a\b\g]$.  The procedure for
fixing these ambiguities is the same as the one used in the
previous section. 

The $M=\a$ component of \eqn{diffalgebra} proceeds in the same
way, except that now also the Lorentz transformation in
\eqn{compocomm} enters (not through a tangent-space rotation, but
through the diffeomorphism \eqn{thetalorentz}), 
\be
\Xi^\a(\e)\Big\vert_{\theta^2} = \bar \theta\,\G^\n\e\; \bar \theta\,
\G^\mu\psi_\n\,\psi_\m{}^{\!\a} +  \ft14 \,\bar\theta\,\G^\n\e\; \hat
\omega^{rs}_\n\,(\G_{rs}\theta)^\a  + \e^\b\,N_\b{}^{\!\a} \,.
\ee
Here we also used the condition of vanishing super-torsion \eqn{torsion}.
The quantity $N_\b{}^{\!\a}$ denotes terms proportional to $\hat F\,
\theta^2$, which are much harder to integrate. They are
controlled by the equation 
\bea
&&\e_2^\b\,\pa_\b N_\gamma{}^{\!\a}\, \e_1^\gamma -\bar\theta\, \G^\m\e_2
(T_\m{}^{\,\n\rho\s\lambda}\e_1)^\a \,\hat F_{\n\rho\s\lambda} -
(1\leftrightarrow 2)= \nonumber\\ 
&&\hspace{1cm}  -\ft1{288}
(\G_{rs}\theta)^\a\,\bar\e_2(\G^{rs}{}_{\!\n\rho\s\lambda} + 24\, 
\d^r_\n\d^s_\rho\G_{\s\lambda})\e_1 \, \hat F^{\n\rho\s\lambda} \,.
\label{rr}
\eea

Leaving these terms aside for the moment, we continue with the
vielbein transformations. 
The knowledge of the $\theta^2$-terms in $\Xi^\m$ (c.f.
\eqn{xi-mu-2}) suffices to evaluate the possible contributions to
$E_\a{}^{\!r}$. The local Lorentz transformations do not
contribute at this order in $\theta$  and one finds that all
contributions cancel. This 
enables one to set (up to ambiguities)
\be
E_\a{}^{\!r}\Big\vert_{\theta^2}=0\,.
\ee
Subsequently one considers $E_\m{}^{\!r}$ and obtains
\be
E_\m{}^{\!r}\Big\vert_{\theta^2} = \bar\theta\,\G^r\Big[ -\ft14
\,\hat\omega_\m{}^{\!st}\G_{st} + T_\m{}^{\!\n\rho\s\lambda}\,\hat 
F_{\n\rho\s\lambda}\Big]\theta \,.
\ee
In order to reconcile the variations with a superspace
diffeomorphism, we had to include a tangent-space
transformation defined by 
\be
\Lambda^{rs}(\e)=  \bar\e\,\G^\m\theta\,\hat\omega_\m^{rs} +
\ft1{144} \bar\theta(\G^{rs\m\n\rho\s} \hat F_{\m\n\rho\s} +24 \,\G_{\m\n}
\hat F^{rs\m\n})\e + {\cal O}(\theta^2)\,.
\ee
At this point we can verify that $\Xi^M\,E_M{}^{\!r}$ remains
field-independent and given by the first
equation of \eqn{tangentdiffs} up to terms of order
$\theta^3$. As we already mentioned the second equation of
\eqn{tangentdiffs} will acquire terms proportional to $\hat
F_{rstu}$. As it turns out the vielbein component $E_\a{}^{\!a}$
is only modified by $\hat F\theta^2$-terms. Denoting these 
by $M_\a{}^{\!a}$, they are subject to the following condition,    
\bea
&&\e^\b\pa_\b M_\a{}^{\!a} -\e^\b\, \pa_\a N_\b{}^{\!\gamma}
\,\d^a_\gamma= \\ 
&& \hspace{1cm} -
\Big[(\bar\e\,\G^\m)_\a (T_\m{}^{\,\n\rho\s\lambda} \theta)^a + \ft1{576}
\bar
\theta (\G^{rs\n\rho\s\lambda} + 24
\,\d^r_\n\d^s_\rho\G_{\s\lambda})\e  \, 
(\G_{rs})^a{}_{\!\a} \Big] \hat F_{\n\rho\s\lambda} \nonumber\,.\quad
\eea

However, neither the explicit form of these $\hat F
\theta^2$-corrections, nor the   
$\theta^2$-contrib\-utions to the supervielbein $E_\mu{}^a$, are
very relevant from the membrane 
point of view, as they do not appear in the supermembrane
action \eqn{supermem}, which depends only on $\varepsilon^{ijk}\,
\Pi_i^A \Pi_j^B\Pi_k^C  
\,B_{CBA}=\varepsilon^{ijk}\,\partial_iZ^M\partial_jZ^N
\partial_kZ^P\, B_{PNM}$ and $g_{ij}=\Pi_i^r\, \Pi_j^s\, \eta_{rs}$.
Therefore we refrained from determining their explicit form at
this order of $\theta$. 

\subsection{The tensor field at order $\theta^2$}
A brief perusal of the algebra involving the tensor gauge
transformations based on \eqn{diff-tensor} reveals the possible
presence of $(\theta^2\e)$- and $(C\, \theta^2\e)$-terms in
$\Xi_{\m\a}$ and $(\theta^2\e\,\psi)$- and
$(C\,\theta^2\e\,\psi)$-terms in  
$\Xi_{\m\n}$. On the other hand no contributions are indicated
for $\Xi_{\a\b}$. However, we did not attempt to work out the
tensor gauge parameters 
from the algebra, but instead proceeded directly to the
variations of the tensor fields. 
{From} the variations of $B_{\a\b\c}$ one finds 
\be
\Xi_{\a\b}(\e)\Big\vert_{\theta^2}= B_{\a\b\g}\Big\vert_{\theta^2}= 0\,,
\ee
up to tensor gauge transformations.

Subsequently one considers the variations to $B_{\m\a\b}$. These
lead to the gauge parameter
\be
\Xi_{\m\a}(\e)\Big\vert_{\theta^2} = \ft16 \bar\theta\,\G^\n \e \,
(\bar\theta\,\G_{\m\n})_\a+ 
\ft16(\theta\,\G^\n)_\a\, \bar\theta\,\G_{\m\n}\e \,,
\ee
and the gauge field components
\be
B_{\m\a\b}\Big\vert_{\theta^2}= (\bar\theta\,\G_{\m\n})_{(\a}\,
(\bar\theta\,\G^\n)_{\b)} \,.
\ee
In obtaining this result we reordered the fermions by making use
of \eqn{4spinor-id}. 
Again these results are not unique and can be changed by a
subsequent tensor gauge transformation with parameters
proportional to $\theta^3$. In this gauge the expression for
$B_{\m\n\a}$ agrees with the flat-space result \eqn{flatssquantities}. 

The variations of $B_{\m\n\a}$ proceed in a similar way and we
find
\bea
\Xi_{\m\n}(\e)\Big\vert_{\theta^2} &=& \bar\theta
\,\G^\rho\e\;\bar\theta(C_{\m\n\s}\,\G^\s +  
\G_{\m\n})\psi_\rho 
+ \ft43 \bar\theta\,\G^\rho\psi_{[\m}\;\bar\theta\,\G_{\n]\rho}\e 
+\ft43
\bar\theta\,\G^\rho\e \;\bar\theta\,\G_{\rho[\m}\psi_{\n]}\,,
\nonumber \\ 
B_{\m\n\a}\Big\vert_{\theta^2} &=&- \ft83 \bar\theta\,\G^\rho
\psi_{[\m}\;(\bar\theta\,\G_{\n]\rho})_\a +\ft43
(\bar\theta\,\G^\rho)_\a \;\bar\theta\,\G_{\rho[\m}\psi_{\n]}
\,.
\eea
Again these results are subject to change under tensor gauge
transformations. We used \eqn{4spinor-id}, just as in the
evaluation of the remaining component, $B_{\m\n\rho}$, which
yields the result
\be
B_{\m\n\rho}\Big\vert_{\theta^2} =
-3\,\bar\theta\,\G_{[\m\n}\Big[-\ft14\hat\omega_{\rho]}{}^{\!rs}\,\G_{rs}
+ T_{\rho]}{}^{\!\s\lambda\kappa\tau}\,\hat
F_{\s\lambda\kappa\tau}\Big]\theta  - 12\,
\bar\theta\,\G_{\s[\m} \psi_{\n}\;\bar\theta \,\G^\s \psi_{\rho]}
\,. 
\ee

\subsection{Summary of the results}

In this subsection we summarize the combined results of this and
the previous section. 
We first present the expressions for the vielbein and the
antisymmetric tensor field. Subsequently we give the expressions
for the superspace transformations in terms of the component
fields and transformation parameters. As the 11-dimensional
coordinate transformations act in the standard way, we only list
the superspace parameters corresponding to supersymmetry and local
Lorentz transformations. 

At order $\theta^2$ we have not fully determined the terms
contributing to $E_\m{}^{\!a}$ and neither did we fully determine
the $\hat F \,\theta^2$-terms in $\Xi^\a$ and
$E_\a{}^{\!a}$. Our results are in agreement with those of
\cite{CF} in corresponding orders of $\theta$. 
While high-rank tensors are of course absent in 4
dimensions, there is a clear similarity between our results and
those in 4 dimensions \cite{FVN}. 
\subsubsection{Vielbein and tensor field expressions}
For the supervielbein $E_M{}^{\!A}$ we found the following
expressions, 
\bea
E_\m{}^{\!r} &=& e_\m{}^r + 2\, \bar\theta\,\G^r \psi_\m
\nonumber \\
&& 
+ \bar\theta\,\G^r\Big[ -\ft14
\,\hat\omega_\m{}^{\!st}\G_{st} + T_\m{}^{\!\n\rho\s\lambda}\,\hat 
F_{\n\rho\s\lambda}\Big]\theta 
+  {\cal O}(\theta^3)\,, \nonumber \\
E_\m{}^{\!a} &=& \psi_\m{}^{\!a} - \ft14\,
\hat\o_\m^{rs}\,(\G_{rs} \theta)^a + 
(T_\m{}^{\!\n\rho\s\lambda}\theta)^a\, \hat F_{\n\rho\s\lambda} +
{\cal O}(\theta^2)\,, \nonumber \\
E_\a{}^{\!r} &=& -(\bar\theta\,\G^r)_\a +  {\cal O}(\theta^3)\,,
\nonumber \\  
E_\a{}^{\!a} &=& \d^a_\a + M_\a{}^{\!a} + {\cal O}(\theta^3)\,,
\label{svielbein}
\eea
where $M_\a{}^{\!a}$ characterizes the $\hat
F\theta^2$-contributions, which we did not evaluate
explicitly. Observe that we determined $E_\m{}^{\!a}$ only up to
terms of order $\theta^2$. 
The result for the tensor field $B_{MNP}$ reads as follows,
\bea
B_{\m\n\rho} &=& C_{\m\n\rho} -6\, \bar \theta
\G_{[\m\n}\psi_{\rho]}  \nonumber\\
&& -3\,\bar\theta\,\G_{[\m\n}\Big[-\ft14\hat\omega_{\rho]}{}^{\!rs}\,\G_{rs}
+ T_{\rho]}{}^{\!\s\lambda\kappa\tau}\,\hat
F_{\s\lambda\kappa\tau}\Big]\theta  - 12\,
\bar\theta\,\G_{\s[\m} \psi_{\n}\;\bar\theta \,\G^\s \psi_{\rho]}
  +  {\cal O}(\theta^3) \,,\nonumber\\
B_{\m\n\a} &=& (\bar \theta \,\G_{\m\n})_\a - \ft83 \bar\theta\,\G^\rho
\psi_{[\m}\;(\bar\theta\,\G_{\n]\rho})_\a +\ft43
(\bar\theta\,\G^\rho)_\a \;\bar\theta\,\G_{\rho[\m}\psi_{\n]}
+  {\cal O}(\theta^3) \,, \nonumber\\ 
B_{\m\a\b} &=& (\bar\theta\,\G_{\m\n})_{(\a}\,
(\bar\theta\,\G^\n)_{\b)} 
+  {\cal
O}(\theta^3) \, ,  \nonumber \\
B_{\a\b\gamma} &=& (\bar\theta\Gamma_{\m\n})_{(\a}\, 
(\bar\theta\Gamma^\m)_{\b}\, 
(\bar\theta\Gamma^\n)_{\g)}
+  {\cal O}(\theta^3) \,. 
\label{stensor}
\eea
For completeness we included the $\theta^3$-term in
$B_{\a\b\gamma}$ which is known from the flat-superspace results.

\subsubsection{Supersymmetry transformations}
The supersymmetry transformations consistent with the fields
specified above, are generated by superspace diffeomorphisms,
local Lorentz transformations and tensor gauge
transformations. The corresponding parameters are as follows. For
the superspace diffeomorphisms are expressed by 
\bea
\Xi^\m(\e) &=& \bar\theta\, \G^\m \e -\bar \theta\,
\G^\n\e\;\bar\theta\,\G^\m\psi_\n
+ {\cal O}(\theta^3)\,, \nonumber\\ 
\Xi^\a(\e) &=& \e^\a -\bar\theta \,\G^\m\e\,\psi_\m^\a \nonumber \\
&&+ \bar \theta\,\G^\n\e\; \bar \theta\,
\G^\mu\psi_\n\,\psi_\m{}^{\!\a} +\ft14   \bar\theta\,\G^\n\e \,
\hat\omega^{rs}_\n\,(\G_{rs}\theta)^\a  + \e^\b \,N_\b{}^{\!\a} 
+  {\cal O}(\theta^3)\,, \label{sdiffs}
\eea 
where $N_\b{}^{\!\a}$ encodes the terms proportional to $\hat F
\theta^2$. The Lorentz transformation is given by
\be
\Lambda^{rs}(\e) =  \bar\e\,\G^\m\theta\,\hat\omega_\m^{rs} +
\ft1{144} \bar\theta(\G^{rs\m\n\rho\s} \hat F_{\m\n\rho\s} +24 \,\G_{\m\n}
\hat F^{rs\m\n})\e + {\cal O}(\theta^2)\,. \label{slorentz}
\ee
Finally, the tensor gauge transformations are parametrized by 
\bea
\Xi_{\m\n}(\e)&=& \bar \e(C_{\m\n\rho}\,\G^\rho + \G_{\m\n}
)\theta 
+\bar\theta\,\G^\rho\e\;\bar\theta(C_{\m\n\s}\,\G^\s +  
\G_{\m\n})\psi_\rho 
+ \ft43 \bar\theta\,\G^\rho\psi_{[\m}\;\bar\theta\,\G_{\n]\rho}\e
\nonumber \\  
&& +\ft43
\bar\theta\,\G^\rho\e \;\bar\theta\,\G_{\rho[\m}\psi_{\n]}
+  {\cal O}(\theta^3)\,, \nonumber \\
\Xi_{\m\a}(\e) &=& \ft16 \bar\theta\,\G^\n \e \,
(\bar\theta\,\G_{\m\n})_\a+ 
\ft16(\theta\,\G^\n)_\a\, \bar\theta\,\G_{\m\n}\e +  {\cal
O}(\theta^3)\,, \nonumber \\ 
\Xi_{\a\b}(\e) &=& {\cal O}(\theta^3) \,.
\label{sgauge}
\eea
\subsubsection{Local Lorentz transformations}
Local Lorentz transformations are generated by a superspace local
Lorentz transformation combined with a diffeomorphism. The
corresponding expressions are given by 
\bea
\Lambda^{rs}(\lambda)&=&\lambda^{rs}\,, \nonumber\\
\Xi^\a(\lambda)&=& -\ft14\lambda^{rs} (\G_{rs}\theta)^\a\,. 
\eea

\section{The supermembrane in background fields}
\setcounter{equation}{0}
The initial supermembrane action \eqn{supermem} is manifestly covariant
under independent superspace diffeomorphisms, tangent-space
Lorentz transformations  and tensor gauge
transformations. For the specific superspace fields associated
with 11-dimensional on-shell supergravity, this is no longer
true and one has to restrict oneself to the superspace
transformations corresponding to the component supersymmetry,
general-coordinate, local Lorentz and tensor gauge 
transformations. When writing \eqn{supermem} in components,
utilizing the expressions found in the previous sections, one
thus obtains an action that is covariant under the restricted
superspace 
diffeomorphisms \eqn{sdiffs} acting on the superspace
coordinates $Z^M=(X^\m,\theta^\a)$ (including the spacetime
arguments of the background fields) combined with usual
transformations on the component fields (we return to this point
shortly). Note that the result 
does not constitute an invariance. Rather it implies that the
actions corresponding to two different 
sets of background fields that are equivalent by a component
gauge transformation, are the same modulo a reparametrization of
the supermembrane embedding coordinates. More precisely, if we
denote the superspace coordinates by $\phi$ and the background by
$G(\phi)$, then the action satisfies
$S[\phi,G]=S[\phi^\prime ,G^\prime]$, where
$G^\prime$ is related to $G$ by a component supersymmetry
transformation. Of course, when
considering a background that is invariant under (a subset of)
the component transformations (so that $G=G^\prime$), then the
action will be invariant 
under the corresponding change of the supercoordinates. 

Using the previous results we may now write down the complete
action of the supermembrane coupled to background fields up to order
$\theta^2$. Direct substitution leads to the following result for
the supervielbein pull-back,
\bea
\Pi_i^r&=&\partial_iX^\mu\, \Bigl (e_\m{}^r + 2\,
\bar\theta\,\G^r \psi_\m 
-\ft14 \bar\theta\,\G^{rst}\theta\,\hat\omega_{\m\,\!st} + \bar
\theta\,\G^r T_\m{}^{\!\n\rho\s\lambda}\theta \,\hat 
F_{\n\rho\s\lambda} \Big) \nonumber \\
&&
+ \bar\theta\Gamma^r\partial_i\theta + {\cal
O}(\theta^3)\,,\nonumber\\
\Pi_i^a&=& \pa_iX^\m \Big(\psi_\m{}^{\!a} - \ft14\,
\hat\o_\m^{rs}\,(\G_{rs} \theta)^a + 
(T_\m{}^{\!\n\rho\s\lambda}\theta)^a\, \hat
F_{\n\rho\s\lambda}\Big)\nonumber \\
&&+ \pa_i\theta^a  +  {\cal O}(\theta^2)\,.
\label{vielbeinpb}
\eea
Consequently the induced metric is known up to terms of order
$\theta^3$. 

Furthermore the pull-back of the tensor field equals 
\bea
\lefteqn{-\ft{1}{6}\varepsilon^{ijk}\,
\Pi_i^A\,\Pi_j^B\,\Pi_k^C\, B_{CBA} =  
 -\ft16\vep^{ijk} \pa_iZ^M\,\pa_jZ^N\,\pa_kZ^P\, B_{PNM} =} \nn\\
&& \ft{1}{6}\, dX^{\mu\nu\rho}\, \Big[ 
C_{\m\n\rho} -6\, \bar \theta
\G_{\m\n}\psi_{\rho}
+\ft34\bar\theta\,\G_{rs}\G_{\m\n}\theta\,\hat\omega_{\rho}{}^{\!rs}
\nn\\
&&\hspace{17mm} - 3 \,\bar\theta\,\G_{\m\n}
T_{\rho}{}^{\!\s\lambda\kappa\tau}\theta \,\hat
F_{\s\lambda\kappa\tau} - 12\,
\bar\theta\,\G_{\s\m} \psi_{\n}\;\bar\theta \,\G^\s
\psi_{\rho}\Big]\nonumber\\ 
&&- \varepsilon^{ijk} \,
\bar\theta\,\Gamma_{\mu\nu}\partial_k\theta \Big
[{\textstyle{1\over 2}} 
\,\partial_i X^\mu (\partial_j X^\nu +
\bar\theta\,\Gamma^\nu\partial_j \theta) + {\textstyle{1\over 6}}
\,\bar\theta\,\Gamma^\mu\partial_i\theta\;
\bar\theta\,\Gamma^\nu\partial_j\theta \Big] \nn\\
&& +\ft13 \varepsilon^{ijk} \pa_i X^\m \,\pa_jX^\n \Big[
4\,\bar\theta \,\G_{\rho\m}\pa_k\theta \;
\bar\theta\,\G^\rho\psi_\n - 2 \,\bar \theta\,
\G^\rho\pa_k\theta\; \bar\theta\,\G_{\rho\m}\psi_\n\Big] 
+ {\cal O}(\theta^3)\,, \quad
\label{WZWpb}
\eea
where we have introduced the abbreviation $dX^{\mu\nu\rho}=
\varepsilon^{ijk}\,  
\partial_i X^\mu\,\partial_j X^\nu\, \partial_k X^\rho$ for the
world-volume form. Observe that we included also the 
terms of higher-order $\theta$-terms that were determined in
previous sections and listed in \eqn{stensor}. We will return to
these terms at the end of this section.
The first formula of \eqn{vielbeinpb} and
\eqn{WZWpb} now determine the supermembrane action \eqn{supermem}
up to order $\theta^3$.

As an illustration of what we stated at the beginning of this
section, we consider the effect of the superspace diffeomorphisms
\eqn{sdiffs} on $\Pi^A_i$. We only need the variations to first
order in $\theta$, so that we substitute
$X^\mu\to X^\m+ \bar\theta\G^\m\e$ and $\theta \to \theta + \e -\bar
\theta\,\G^\m\e\;\psi_\m$  into \eqn{vielbeinpb}. For $\Pi^r_i$
this induces a variation which can be rewritten as 
\be
\d\Pi^r_i = \pa_i X^\m \Big[\d e_\m{}^{\!r} +
2\bar\theta\,\G^r\d\psi_\m \Big] - \Lambda^{rs}(\e)
\,\Pi^s_i + {\cal O}(\theta^2)\,.
\ee
The first term on the right-hand side represents the change of
$\Pi^r_i$ under the supersymmetry variations \eqn{susycomp} of
the background fields. The second term represents a Lorentz
transformation whose parameter is given by
\eqn{slorentz}. For the induced metric, given by $g_{ij}=
\Pi_i^r\,\Pi^s_j\,\eta_{rs}$, the Lorentz transformation drops
out, so that the effect of the coordinate change of
$(X^\m,\theta^\a)$ is the same as when performing a supersymmetry
transformation of the background 
fields. This implies that the 
first term in the supermembrane action \eqn{supermem} has indeed
the required transformation behaviour.

A similar result holds for the variation of $\Pi_i^a$ under the
coordinate change, but only in zeroth-order in $\theta$, as we
have not determined all the $\theta^2$-contributions. An explicit
calculation gives, 
\be
\d\Pi_i^a= \pa_iX^\m\Big[ \d\psi_\m^a - \bar\theta\,\Gamma^\n\e\;
\hat \psi_{\m\n} \Big] + \bar\theta\,\Gamma^\m
\e\;\Big((\ft14\hat\omega_\m^{rs} 
\G_{rs} - T_\m{}^{\!\n\rho\s\lambda}\hat
F_{\n\rho\s\lambda})\,\Pi_i\Big)^a\,,
\ee
where $\hat\psi_{\m\n}$ is the supercovariant curl of the gravitino
field. As expected, the terms linear in $\theta$ do not exhibit
the same systematics. But we do not need the expression for
$\Pi^a_i$ for the supermembrane action, so that this issue is not
of immediate relevance. 

Let us now consider the variation of the second term \eqn{WZWpb}
in the supermembrane action \eqn{supermem}. Its variation takes
the form 
\bea
\d\Big(\!-\ft{1}{6}\varepsilon^{ijk}\,
\Pi_i^A\,\Pi_j^B\,\Pi_k^C\, B_{CBA}\Big)&\!\!=\!\!& 
-\pa_i\Big[\, \ft12\varepsilon^{ijk}
\,\pa_jX^\m\,\pa_kX^\n\,\bar\e(C_{\m\n\rho}\G^\rho +\G_{\m\n})
\theta\,\Big] \nn\\
&& 
+ \ft16dX^{\m\n\rho} \Big[ \d C_{\m\n\rho} -6\, \bar\theta\, 
\d(\G_{\m\n}\psi_\rho) \Big] + {\cal O}(\theta^2) \,. \quad
\label{covWZW}
\eea
To show that all explicit $\psi^2$-terms cancel, we again made 
use of \eqn{4spinor-id}. The above results shows that also for the
second term of the supermembrane action, a supersymmetry
coordinate change gives the same effect as a supersymmetry
transformation of the background fields (up to a
world-volume surface term which we will discuss in more detail at
the end of this section). 

While the above results were guaranteed to hold on the basis of
the procedure followed in sections~3 and 4, the next feature is
independent of that and concerns the $\kappa$-invariance of the
action. The $\kappa$-symmetry transformations are defined in the
unrestricted superspace and will be given below. In principle, it
should be possible to derive the transformation rules in the
gauge-fixed superspace situation that we are working
with. However, it is not necessary to do so, because we
are only interested in establishing the invariance of the
action. Both the original and the gauge-fixed action should be
$\kappa$-symmetric, so that we can just use the original
superspace diffeomorphisms corresponding to $\kappa$-symmetry and
substitute them in the gauge-fixed 
action. These $\kappa$-transformations take the form of superspace
coordinate changes defined by \cite{BST}
\be
\delta Z^M\, E_M{}^r = 0 \,,\qquad \delta Z^M\, E_M{}^a =
(1-\Gamma)^a{}_b\, \kappa^b \,, \label{kappatransf}
\ee
where $\kappa^a(\zeta)$ is a local fermionic parameter and the
matrix $\Gamma$ is defined by
\be
\Gamma= \frac{\varepsilon^{ijk}}{6\sqrt{-g}}\, \Pi^r_i
\,\Pi^s_j\, \Pi^t_k\, \Gamma_{rst}  \, ,
\label{supergamma}
\ee
with $g=\det g_{ij}$. It satisfies the following properties,
\be
\Gamma^2=1 \,,\qquad \Gamma\,\G_r \Pi^r_i =  \Pi^r_i
\G_r\,\Gamma =  \ft12 {g_{ij}\over
\sqrt{-g}}\,\varepsilon^{jkl}\, \Pi^r_k\,\Pi^s_l\, \G_{rs}
\,. \label{gammaproperties} 
\ee   
Therefore the matrix $(1-\Gamma)$ 
in \eqn{kappatransf} is a projection operator. As a
consequence, this allows one to gauge away half of the $\theta$
degrees of freedom. 

It is advantageous to expand the $\kappa$-transformations
\eqn{kappatransf} as follows,   
\bea
\delta X^\mu&=& \bar\kappa_- \Gamma^\mu \theta
- \bar\kappa_-\Gamma^\n\theta\; \bar\theta\,\Gamma^\mu\psi_\n
+ {\cal O}(\theta^3) \,,\nn \\
\delta \theta &=&  \kappa_- 
+ \psi_\mu \;\bar\theta\,\Gamma^\mu\kappa_-  + {\cal O}(\theta^2)\,,
\label{dktheta}
\eea
where we have introduced the chiral spinor
$\kappa_-=(1-\Gamma)\kappa$. We stress that we are not making any
approximation in $\G$, which depends on the background fields and
on $\theta$ in a complicated fashion. Note that we retain the
$\theta^2$-contributions to $\delta X^\m$ for reasons that will
become clear shortly. Under the variations
\eqn{dktheta} we then derive the following result, 
\bea
\d\Pi^r_i&=& 2\,\bar\km \,\G^r\pa_i\theta\nn\\
&&
+ \Pi^s_i\Big[ 2\,\bar\km \, \G^r \psi_s - 4 \,\bar
\theta\,\G^\m \psi_s \;\bar\km \, \G^r\psi_\m -\ft12 \bar\km
\,\G^r \G_{tu} \theta \,\hat\omega_s{}^{\!tu} + \bar\km
\,\G^\n\theta \,\hat\omega_\n{}^{\!r}{}_s  \nn\\
&&
\hspace{10mm} + \bar\km ( \G_s T^{r\,\n\rho\s\lambda} + \G^r
T_s{}^{\!\n\rho\s\lambda} + 4e^{r\n}
\G_s{}^{\!\rho\s\lambda}-4e_s{}^{\!\n} \,\G^{r\rho\s\lambda} )
\theta\,\hat F_{\n\rho\s\lambda} \Big] \nn\\
&&
+ {\cal O}(\theta^2) \,. 
\eea
Here we rewrote the right-hand side in terms of
$\Pi^s_i$, rather than $\pa_i X^\m$. This is the origin of the
explicit $\psi^2$-term; all other explicit $\psi^2$-terms cancel. 
It is now straightforward to obtain the $\kappa$-variation of the
induced metric, 
\bea
\d g_{ij} &\!\!=\!\!& 4\,\Pi^r_{(i} \;\bar\km
\,\G^r\pa_{j)}\theta\nn\\ 
&&
+ \Pi^r_{(i}\,\Pi^s_{j)}\, \Big[ 4\,\bar\km \, \G_r \psi_s - 8 \,\bar
\theta\,\G^\m \psi_s \;\bar\km \, \G_r\psi_\m \nn\\
&&
\hspace{18mm} - \bar\km
\,\G_r \G_{tu} \theta \,\hat\omega_s{}^{\!tu} +4 \, \bar\km \, \G_r
T_s{}^{\!\n\rho\s\lambda} 
\theta\,\hat F_{\n\rho\s\lambda} \Big] + {\cal O}(\theta^2) \,.\quad
\eea

Subsequently we consider the variation of the second term of the
supermembrane action\eqn{supermem}, which yields 
\bea
\lefteqn{\d\Big(\!-\ft{1}{6}\varepsilon^{ijk}\,
\Pi_i^A\,\Pi_j^B\,\Pi_k^C\, B_{CBA}\Big) =
-\varepsilon^{ijk}\, \Pi_i^r \,\Pi_j^s
\;\bar\km\,\G_{rs}\pa_k \theta }  \nn\\
&& + \varepsilon^{ijk} \,\Pi^r_i\,\Pi^s_j\,\Pi^t_k\Big[ -
\bar\km\, \G_{rs}\psi_t  + 2\,\bar\theta\,\G^\mu \psi_t
\;\bar\km\,\G_{rs}  \psi_\m    \nn\\
&&\hspace{29mm} 
+\ft14 \bar\km\,\G_{rs}\G_{uv} \theta\,\hat\omega 
^{uv}_t - \bar\km\,\G_{rs} T_t{}^{\!\n\rho\s\lambda} \theta
\,\hat F_{\n\rho\s\lambda} \Big] + {\cal O}(\theta^2) \, \quad
\label{varwzw}
\eea
up to a total derivative which we will discuss shortly. 
In deriving this result, we again used \eqn{4spinor-id} to reorder
the $\psi^2$-terms.  

At this point it is rather easy to establish the $\kappa$-invariance of
the action. Observing that the variation of the first term in the
action is
equal to $- \d \sqrt{-g}= -\ft12 \sqrt{-g} \,g^{ij}\,\d g_{ij}$,
replacing $\bar\km$ by $-\bar\km\,\G$ and making use
of the second equation \eqn{gammaproperties}, one verifies 
directly that the variations of the two terms in the Lagrangian
\eqn{supermem} vanish under $\kappa$-symmetry, up to a surface
term.

Hence we have verified that up to first order in $\theta$
the supermembrane action
transforms as a scalar under supersymmetry and is invariant under
$\kappa$-symmetry, up to a world-volume surface term. 
Let us stress that at this order no need arose to make use of the
11-dimensional supergravity field equations in verifying the
$\kappa$-symmetry of the action. We expect that this will be necessary
at higher orders as is indicated by the analysis of \cite{BST}.
In order to check these symmetries at second order we would have
to know the supermembrane action up to third order in $\theta$, as
the supersymmetry as well as the $\kappa$-transformation of $\theta$
contain terms of zeroth order in $\theta$.  

Finally, let us return to the surface terms that did not receive
much attention earlier when establishing the supersymmetry and the
$\kappa$-invariance. These terms are relevant when considering
the open supermembrane \cite{openmb,open1,open2,Astrings}. As it
turns out, we can easily determine the surface term, including
some contributions of higher order in $\theta$. 
For $\kappa$-symmetry, we observe that all
variations proportional to $\pa_i\kappa_-$ must be generated by
the surface term. Assuming that $\kappa$-symmetry is valid, the
surface contributions can therefore be evaluated by simply
collecting all variations proportional to world-volume
derivatives of $\kappa_-$.  
Moreover, these terms can only come from the Wess-Zumino-Witten
sector, because the pull-back $\Pi^r_i$ does not
generate derivatives of $\kappa_-$ owing to the first equation
of \eqn{kappatransf}. 

For supersymmetry, $\Pi^r_i$ and the Wess-Zumino-Witten term are
separately invariant in the sense explained earlier and surface
terms can only come from the latter. One can thus use the same
strategy and collect the variations proportional to  derivatives
of the supersymmetry parameter. 
However, now these terms come from two sources, namely from 
$\pa_iZ^M$ and from the gravitino terms. To see how this works,
one may compare to the calculation leading to \eqn{covWZW}.

Because we know the variations  $\d X^\m$ to second and $\d\theta$
to first order in $\theta$ for both supersymmetry and
$\kappa$-symmetry, we can determine the surface
contributions to order ${\rm d}X\wedge{\rm d}X\,\theta^2$
and ${\rm d}X\wedge{\rm d}\theta\,\theta^2$, while, for
$\kappa$-symmetry, we also reliably
calculate those terms of the form ${\rm d}\theta\wedge{\rm
d}\theta \,\theta^3$ which are present in the flat-superspace
case. In this way we find the following results. The surface term
associated with $\kappa$-symmetry is given by
\bea
&&\int_{\partial M} \,\bigg[\,{\rm d}X^\mu \wedge {\rm d}X^\nu \,\Big[ -\ft12
\bar \theta(C_{\m\n\rho}\G^\rho + \G_{\m\n}) \kappa_- - 
\ft12 \bar\psi_\s(C_{\m\n\rho}\G^\rho + \G_{\m\n})\theta\;
\bar\theta \,\G^\s \kappa_- \nn\\ 
&& \hspace{35mm} 
+\ft43 \bar\psi_\m( \G^\rho \theta\;\bar\theta\, \G_{\rho\n} + 
\G_{\rho\n} \theta\;\bar\theta\, \G^\rho )\kappa_-\Big]  \nn\\ 
&&\hspace{11mm}+ \ft12 {\rm d}X^\mu \wedge {\rm d}\bar\theta
\,\Big[  \G^\n \theta\;\bar\theta\, \G_{\m\n} +  
\G_{\m\n} \theta\;\bar\theta\, \G^\n \Big]\kappa_- \nn\\  
&&\hspace{11mm} + \ft16 \bar\theta\,\G^\m  {\rm d}\theta \wedge {\rm
d}\bar\theta \,\Big[ \G^\n 
\theta\;\bar\theta\, \G_{\m\n} +
\G_{\m\n} \theta\;\bar\theta\, \G^\n+ \cdots \Big]  \kappa_-  \bigg]\,.
\eea
After taking into account the gravitino variations as described
above, the surface term associated with supersymmetry can be
evaluated and equals 
\bea
\label{susyboundary}
&&\int_{\partial M} \,\bigg[\,{\rm d}X^\mu \wedge {\rm d}X^\nu \,\Big[ \ft12
\bar \theta(C_{\m\n\rho}\G^\rho + \G_{\m\n}) \e +
\ft12 \bar\psi_\s(C_{\m\n\rho}\G^\rho + \G_{\m\n})\theta\;
\bar\theta \,\G^\s \e \nn\\ 
&& \hspace{35mm} 
- \ft23 \bar\psi_\m( \G^\rho \theta\;\bar\theta\, \G_{\rho\n} + 
\G_{\rho\n} \theta\;\bar\theta\, \G^\rho )\e\,\Big]  \nn\\ 
&&\hspace{11mm}- \ft16 {\rm d}X^\mu \wedge {\rm d}\bar\theta
\,\Big[  \G^\n \theta\;\bar\theta\, \G_{\m\n} +  
\G_{\m\n} \theta\;\bar\theta\, \G^\n \Big]\e +\cdots \bigg] \,.
\eea
Here the terms proportional to ${\rm d}\theta\wedge {\rm
d}\theta$ cannot be determined, because the corresponding
gravitino terms have not been obtained to sufficiently high order of
$\theta$. The background-independent terms in \eqn{susyboundary} coincide with those
given by \cite{open2} in the flat-superspace case. This provides
another nontrivial verification of the correctness of our
results. 

The difference with the corresponding flat-case expressions
\cite{open1,open2,Astrings} resides in the coupling to
$C_{\mu\nu\rho}$ and $\psi_\mu$. However, most of the surface
terms cancel by assuming a
``membrane D-$p$-brane'' at the boundary and imposing the
Dirichlet conditions 
\be
\partial_\parallel X^{\tilde m}  = 0\,,  \quad \mbox{for } \tilde
m=p+1,\ldots,10, 
\ee
where $\partial_\parallel$ defines the world-volume derivative
tangential to the boundary surface. For the fermionic quantities
$\theta, \psi_m, \e$ and $\kappa_-$ one can impose a projection
such that the only nonzero fermionic bilinears 
involve a product of an odd number of gamma matrices $\G^m$,
where $m= 0,1,\ldots, p$. This requires $p$ to take the values $1,5$
or $9$ \cite{open2,Astrings}. One is thus left with terms proportional to $C_{mnq}$
living on the $p$-brane at the boundary, which
can presumably be dealt with by a deformation of these fermionic
conditions \cite{open2}. Of course, these terms are 
subject to the tensor gauge transformations of 11-dimensional
supergravity. 
This issue can be resolved by having additional
degrees of freedom at the boundary of the membrane. In this
connection it is relevant to observe that the
11-dimensional supergravity action itself is also not invariant
under tensor gauge transformations in the presence of a
boundary. Some of this has been discussed, for instance, in
\cite{horvw,open1}.

\section{Discussion}

In this paper we constructed the superspace vielbein and the
tensor gauge field of 11-dimensional on-shell supergravity in terms
of its component fields to higher orders in $\theta$
coordinates. This enabled 
us to write down the 11-dimensional supermembrane action coupled to 
a nontrivial supergravity component-field background to second
order in $\theta$. 
We then displayed its transformation properties under supersymmetry
and exhibited the invariance of the supermembrane action under the local
fermionic $\kappa$-symmetry, yielding an independent check of
our superspace results. Furthermore we obtained the leading
background-dependent terms of the surface terms for open
supermembranes. 

Having this explicit form of the supermembrane action at ones
disposal  now opens up a 
multitude of interesting applications. The most prominent next step 
is the study of the supermembrane degrees of freedom in background
geometries. In analogy to the
bosonic case discussed in this paper, the light-cone supermembrane 
turns out to be equivalent to a gauge theory of area-preserving 
diffeomorphisms coupled to background fields, modulo
corresponding assumptions on the background geometry.
This U($\infty$) gauge theory
may then in turn be regularized by a supersymmetric
U($N$) quantum-mechanical model in curved backgrounds. Whether or
not this will shed some light on the problem of formulating
matrix models in curved spacetime remains to be seen. A 
conceptually better posed problem concerns perhaps the membrane and the
matrix  models in a constant antisymmetric tensor
background. Other investigations of the supermembrane will deal
with specific background solutions with a certain amount of residual
supersymmetry. 
Interesting candidates for such backgrounds are the
membrane \cite{sugramem} and the fivebrane solution \cite{sugrafive}
of 11-dimensional supergravity,
as well as solutions corresponding to the product of Anti-de-Sitter 
spacetimes with compact manifolds \cite{BDPS}. 
Coupling to AdS solutions appears especially
appealing in view of the recent results on the duality of  
large-$N$ superconformal field theories and supergravity on a
product of  AdS space with a compact manifold \cite{maldacena}.

\bigskip
\vspace{6mm}
\noindent
{\bf Acknowledgements}\\
We thank Jan de Boer, Michael Douglas, Michael Green, Hermann Nicolai, Hirosi
Ooguri, Jan-Willem van Holten and Andrew Waldron for valuable
comments and discussions. 
This research was supported in part by the National Science Foundation
under Grant No. PHY94-07194 through the Institute for Theoretical
Physics in Santa Barbara. B.~de~Wit thanks the Institute for
the hospitality extended to him during this work. 

\vfill

\end{document}